\documentstyle[pra,aps,multicol]{revtex}
\begin{document}

\draft
\title{Edge Modes in the Hierarchical Fractional Quantum Hall Liquids
with Coulomb Interaction}
\author{Takao Morinari and Naoto Nagaosa}
\address{Department of Applied Physics, University of Tokyo,
Bunkyo-ku, Tokyo 113, Japan}

\date{\today}
\maketitle
\begin{abstract}
 Edge modes are studied for the hierarchical  
fractional quantum Hall liquids (FQHL) by treating the edge and bulk in a
unified fashion. Within the RPA treatment of the composite boson effective 
theory with Coulomb interaction, one edge magneto-plasmon is shown to be 
the eigen-mode of the system. 
This mode decays algebraically perpendicular to the edge.
All the other modes decay faster than power-law, and correspond
to the neutral modes. However these neutral modes generally couple with the 
electromagnetic field within the magnetic length from the edge.
The time-of-flight experiment in the $\nu=2/3$ FQHL is 
discussed in terms of these results.

\end{abstract}
\pacs{ }

\multicols{2}
%narrowtext

The fractional quantum Hall liquids(FQHL) are the incompressible quantum 
fluids, and the density fluctuation costs finite energy.
Then the low lying excitations are possible only along the edge of the sample,
which constitute the edge modes.
These one-dimensional edge modes are described as the 
chiral-Tomonaga-Luttinger (CTL) liquids \cite{wen} .
Recently, Milliken et al. observed at the Landau level filling factor 
$\nu=1/3$, that the conductance $G$ is proportional to 
$T^{2/\nu-2}$ which coincides with the prediction by Kane and Fisher 
about Tomonaga-Luttinger liquid \cite{milliken} .
Then it seems that the CTL theory works for 
the filling fraction $\nu=1/(2k+1)$ ($k$: integer).

 At the filling fraction $\nu=m/(mp+1)$ ($m$:integer, $p$:even), 
it is believed that FQHS has $|m|$ CTL modes. 
Especially in the case $m<0$, it is predicted that there is one edge mode which
has chirality opposite to the other modes.
 
 However the time of flight experiment of the pulse injected into the 
$\nu=2/3$ FQHL has detected only one chirality corresponding to the 
electron \cite{ashoori} .

 Kane et al. claimed that the backward scatterings 
caused by the impurities along the edge play a crucial role
for the quantization of the Hall conductance \cite{kanefisher}.
These backward scatterings make the neutral mode massive, 
and only the charged mode carries the Hall current.
 However the fractional quantum Hall effect (FQHE) is a bulk phenomenon, 
and recently one of the present authors have shown 
that the Hall conductance is quantized {\it without } any randomness 
at the edge\cite{nagaosa}.
In this formalism, the edge and bulk are treated in a unified fashion and
the edge modes are derived as the eigen-modes of the 
total system which are localized near the edge.
With the short-range potential, the length scale perpendicular to the edge is
of the order of the magnetic length $\ell_B=\sqrt{c\hbar /eB}$.
The edge currents, which are localized within this length, are
driven by the electric field along the currents while the bulk currents 
are transverse. 

 On the other hand, a theory for the edge magneto-plasmon has been developed
by Volkov and Mikhailov \cite{volkov}. This theory takes 
into account the long-range
Coulomb interaction but the relation between current density $\vec J(r)$ 
and the electric field $\vec E(r)$ is assumed to be that of the bulk up to the
edge of the sample, i.e.,
\begin{equation}
J_{\alpha}(r) = \varepsilon_{\alpha \beta} \sigma_{xy} E_{\beta}(r).
\end{equation}
where $\alpha, \beta = x,y$ and $\varepsilon_{\alpha \beta}= \pm 1$ is the
antisymmetric tensor.
Then this theory does not take into account the edge currents, and 
is considered to be relevant to the length scale much larger
than $\ell_B$, e.g., $ \sim \mu$m.

Then the relationship between these two theories is not clear
at present. Recently Orgad and Levit \cite{orgad} 
have developed a theory for the 
edge mode for the FQHL with $\nu = 1/(2k + 1)$. 
They take into account the long-range Coulomb interaction, which
is similar in spirit to ref. \cite{nagaosa} in the sense that 
the edge mode is derived from the bulk action. 
Orgad and Levit solved the RPA equations with the use of 
the ground state solution which is parametrized 
by the total charge of the system. 
They could treat also the case of the smooth change in the electron 
density near the edge. 
However their treatment is restricted to the case of 
$\nu=1/(2k+1)$, i.e., the single mode case.
On the other hand, Wen takes into account the long range nature of the 
Coulomb interaction in his one-dimensional action \cite{wenreview}.
He concluded that only the magneto-plasmon is charged while all the others are
neutral. However his treatment of the Coulomb interaction is rather
phenomenological, and the spatial dependence perpendicular to the edge
could not be studied. 

In this paper we study the edge modes in the hierarchical FQHL
in terms of the bulk action. This treatment naturally interpolates between the
theory of magneto-plasmon by Volkov-Mikhailov  \cite{volkov} 
and that of the edge modes by Wen \cite{wen}.
This work corresponds to the generalization of the method developed by
Orgad and Levit \cite{orgad} to the hierarchical case
with the filling factor $\nu=m/(m p+1)$ 
( $m$ :integer, $p$ :even).
It is found that only one mode is decaying algebraically and
has the dispersion like $\omega_k \sim  - k \ln k$. 
This is the edge magneto-plasmon, and corresponds to the fluctuation of the 
total charge near the edge.
All the others decay faster than power-law, probably exponential decaying
within the magnetic length $\ell_B$, and appear neutral in the 
longer length scale. 
The hydrodynamic argument reveals that the dispersion of these modes are
$\omega_k \propto k$. Then our microscopic study confirms the 
conclusions by Wen clarifying its range of validity.
From these results, we also discuss the time-of-flight 
experiment of the injected charge to the $\nu=2/3$ FQHL. 

In the case $\nu=m/(mp+1)$, we begin with the boson
Chern-Simons Ginzburg-Landau theory. To begin with, we transform  electrons
to the composite fermions with $p$-flux of the Chern-Simons gauge field 
$a_{\mu}$ attached. Then the electron problem becomes that of the
integer quantum Hall system with $\left| m \right| $-filled Landau levels
for composite fermions \cite{blokwen} .
Each composite fermion is replaced by a boson with a flux
$\phi_0=ch/e$ attached to it. This flux corresponds to the other 
Chern-Simons gauge fields $a_{I\mu}$ with $I=1,2,\cdots,\left| m \right| $.
By integrating over  $a_{\mu}$ after shifting $a_{I\mu}$ to $a_{I\mu}-a_{\mu}$ 
, we obtain the Lagrangian density as
\endmulticols
\vspace{-5mm}\noindent\underline{\hspace{87mm}}
\begin{eqnarray}
{\cal L} & = & \sum_{I=1}^{\left| m \right| } \left[ \rho_I
             \left( -\hbar \partial_t \theta_I - U + e a_{I0} \right)
    -\frac{\rho_I}{2M_I} \left( \hbar \nabla \theta_I + \frac{e}{c} {\bf A}
                + \frac{e}{c} {\bf a}_I \right)^2 \right.
               \left. -\frac{\hbar^2}{2M_I} 
          \left( \nabla \rho_I^{\frac{1}{2}} \right)^2
         + \lambda_I \left( \frac{\rho_b}{\left| m \right|} \rho_I - \frac{1}{2} \rho_I^2 \right)
           \right] \nonumber \\
         &   & -\frac{1}{2} \int d^2 {\bf r}^{\prime} V \left( {\bf r}-{\bf r}^{\prime} \right) 
        \left[ \rho \left( {\bf r},t \right) - \rho_b \left( {\bf r} \right) \right]
       \left[ \rho \left( {\bf r}^{\prime},t \right) - \rho_b \left( {\bf r}^{\prime} \right) \right]
       + \frac{e}{2\phi_0} \nu^{\prime} \sum_{I,J}
   K_{IJ} \epsilon^{\mu \nu \lambda} a_{I\mu} \partial_{\nu} a_{J\lambda}.
\label{lagrangian}
\end{eqnarray}
\noindent
Where $U$ is an external potential, 
$K_{IJ}=p-\left( p \left| m \right| -1 \right) \delta_{IJ}$,
$\nu^{\prime}=1/ (\left| m \right| p -1)$, $\rho = \sum_{I} \rho_I$, and 
$\rho_b$ is the background positive charge density. 
The magnetic field $B$ is in the $-z$ direction.
Each boson field is characterized by the mass $M_I$ and 
the short range hard core interaction $\lambda_I$.
Now we generalize the treatment by Orgad and Levit\cite{orgad} to
derive the edge modes from eq.(\ref{lagrangian}).
Taking the variational derivative with respect to $\rho$, $\theta_I$,
and $a_{I \mu}$, the mean field equations are obtained.
Putting ${\bf v}_I= \left( \hbar \nabla 
\theta_I +e/c \hspace{.2em} {\bf A}+ e/c \hspace{.2em} {\bf a}_I \right)/M_I $,
these mean field equations are
\begin{eqnarray}
 -\hbar \partial_t \theta_I - U + e a_{I0} 
  - \frac{M_I}{2} {\bf v}_I^2
  + \frac{\hbar^2}{2M_I} \frac{1}{\sqrt{\rho_I}} \nabla^2 \sqrt{\rho_I}
% \nonumber \\
%  & & \hspace{.3in} + \lambda_I \left( 
  + \lambda_I \left( 
     \frac{\rho_b}{\left| m \right|} - \rho_I \right)
  - \int d^2 {\bf r}^{\prime} V ( {\bf r} - {\bf r}^{\prime} )
\left[ \rho ( {\bf r}^{\prime} ,t ) - \rho_b ( {\bf r}^{\prime} ) \right] = 0 ,
\label{meaneq1}
\end{eqnarray}
\noindent
\vspace{-5mm}
\begin{equation}
\rho_I = \sum_J K_{IJ} \left( 
   \frac{\nu^{\prime}}{h} M_J \nabla \times {\bf v}_J
                           + \nu^{\prime} \frac{B}{\phi_0} \right) ,
\label{meaneq2}
\end{equation}
\noindent
\vspace{-5mm}
\begin{equation}
\rho_I {\bf v}_I \times \hat{e}_z 
 - \frac{c \nu^{\prime}}{h} \sum_J K_{IJ} 
 \left[
   \partial_t \left( 
  M_J {\bf v}_J - \hbar \nabla \theta_J - \frac{e}{c} {\bf A} \right)
   + \frac{e}{c} \nabla a_{J0} 
 \right] = 0 .
\label{meaneq3}
\end{equation}
\noindent
The equation of continuity is derived from these equations.
The ground state is obtained by solving these equations, but its
explicit form is not needed in the following consideration.

We consider a sample which spreads over the semi-infinite plane($x\geq0$). 
In this geometry, we have the translational invariance along $y$-direction,
then $\rho_I$ and $v_{Iy}$ are the functions of $x$ only and $v_{Ix}=0$.
We choose the Landau gauge for the gauge fields $a_{I\mu}$, i.e. 
${\bf a}_I=(0,a_{Iy}(x))$ 
and ${\bf A}=(0,-Bx)$. 
In this gauge choice, we see that $\partial_y \theta_I$ and
$\partial_t \theta_I$ are functions of $x$ only and 
$\partial_x \theta_I \propto v_{Ix}$ is equal to $0$.
Then we can write $\theta_I = q_I^{\prime} y - (\mu_I/ \hbar) t$
with constant $q_I^{\prime}$ and $\mu_I$.

From (\ref{meaneq2}), the excess charge $q_I \left( =\int_0^{\infty} dx \left( 
\rho_b / \left| m \right| - \rho_I \right) \right)$ of $I$ component 
boson field is related to 
$q_I^{\prime}$ by $q_I = (\nu^{\prime} / 2\pi)
\sum_J K_{IJ} q_J^{\prime}$.
The boson fields equilibrate each other, and the chemical potential for
$I$th component boson is equal to each other, i.e., 
$\mu_I = \mu$ ($I$-independent).

Linearizing the mean field equations around the ground state solutions, we get
the RPA equations :
\begin{equation}
\delta \rho_I - \frac{\nu^{\prime}}{h} \sum_{J} K_{IJ} M_J 
 \left( \partial_x \delta v_{Jy} - \partial_y \delta v_{Jx} \right) =0 ,
\label{rpaeq1}
\end{equation}
\noindent
\vspace{-5mm}
\begin{equation}
\rho_I \delta v_{Iy} + v_{Iy} \delta \rho_I - \frac{c\nu^{\prime}}{h}
 \sum_J K_{IJ} \left[ M_J \partial_t \delta v_{Jx} + \frac{e}{c} \partial_x 
   \delta a_{J0} \right] = 0 ,
\label{rpaeq2a}
\end{equation}
\noindent
\vspace{-5mm}
\begin{equation}
\rho_I \delta v_{Ix} + \frac{c\nu^{\prime}}{h} \sum_J K_{IJ} 
\left[ M_J \partial_t \delta v_{Jy} - \hbar \partial_t \partial_y \delta 
 \theta_J + \frac{e}{c} \partial_y \delta a_{J0} \right] = 0 ,
\label{rpaeq2b}
\end{equation}
\noindent
\vspace{-5mm}
\begin{eqnarray}
\hbar \partial_t \delta \theta_I  =  -M_I v_{Iy} \delta v_{Iy}
 + f \left( \rho_I,\delta \rho_I \right) + e \delta a_{I0} 
           -\lambda_I \delta \rho_I 
                            - \int d^2 {\bf r}^{\prime}
 V \left( {\bf r}-{\bf r}^{\prime} \right) \delta \rho \left( {\bf r}^{\prime},t \right) .
\label{rpaeq3}
\end{eqnarray}
\noindent
%\noindent\hspace{90mm}\underline{\hspace{90mm}}\vspace{-3mm}
\multicols{2}\noindent
where $f$ is given by 
$f \left( \rho,\delta \rho \right) = -(\hbar^2 / 4M_I) \rho^{-3/2} \left(  \nabla^2 \rho^{1/2} \right)$
$ \times \delta \rho + (\hbar^2 / 4M_I) \rho^{-1/2}\nabla^2\left( \rho^{-1/2} \delta \rho \right)$ . We will show below that one solution to these equations is obtained from the ground state configuration.
For this purpose, 
we consider the ground-state solution with the fixed total charge
, i.e., $q=\sum_I q_I$.
We differentiate this ground state configuration 
$\rho_I$ ,$v_{Iy}$, $a_{I0}$ and
$\theta_I$  with respect to the 
total charge $q$, and the derivatives are denoted by 
$\delta \rho_I^{(s)}$ etc. 
When taking the derivative 
we choose $dq_I=dq / \left| m \right|$, which means that
the charge of each component oscillates in phase. 
By using these, we try the solution to eqs.(6)-(9) 
which has the following form.
\begin{equation}
{ \everymath{\displaystyle}
\left\{ \begin{array}{l}
   \delta \rho_I = \delta \rho_I^{(s)}(x) 
   \cos \left( ky-\omega t \right) , \\
   \delta a_{I0} =\delta a_{I0}^{(s)}(x) \cos \left( ky-\omega t \right) , \\
   \delta v_{Iy} =\delta v_{Iy}^{(s)}(x) 
   \cos \left( ky-\omega t \right) , \\
   \delta \theta_I = \frac{2\pi}{\nu^{\prime}} \frac{1}{k} 
              \sin \left( ky-\omega t \right) . \\
\end{array} \right.
}
\label{solutionofrpaeq}
\end{equation}
\noindent
From eq. (\ref{rpaeq2b}), it is seen that 
$\delta v_{Ix}$ is of the order of $k$ or $\omega$ and we put $\delta v_{Ix}=0$.
 We put eq.(\ref{solutionofrpaeq}) and $\delta v_{Ix}=0$ 
into the lhs of eqs.(\ref{rpaeq1}) - (\ref{rpaeq2b}),
and keep only the lowest order terms in $k$ and $\omega$.
After that we obtain the very equations for $\delta \rho^{(s)}$ etc
that coincide with the equations obtained by differentiating 
the ground-state equations by the total-charge $q$. 
Then it has been shown that $\delta \rho_I$ etc. in eq.(\ref{solutionofrpaeq})
satisfy the eqs.(\ref{rpaeq1})-(\ref{rpaeq2b}). 
The last equation (\ref{rpaeq3}) gives the dispersion $\omega_k$ of 
this mode.
In the last term in the 
rhs of eq.(9), the Coulomb interaction leads to the modified Bessel function 
$K_0 ( \left| k \right| x )$.
This function can be approximated as
$K_0 ( \left| k \right| x ) \cong
\log \left( 2e^{-\gamma} / \left| k \right| x \right)$ 
($\gamma$ is the Euler constant),
because the ground state charge density is localized 
in the region of the order of magnetic length $\ell_B$ perpendicular to 
the edge. Then the dispersion is 
\begin{equation}
\hbar \omega_k = 
\frac{\nu}{2\pi} \left( \frac{1}{\left| m \right|} 
\sum_{J=1}^{\left| m \right|}
 \frac{\partial \mu}{\partial q_J} \right) k - \nu \frac{e^2}{\pi} 
 k \log \frac{2e^{-\gamma}}{\left| k \right|}.
\label{dispersion}
\end{equation}
\noindent
From eqs.(\ref{meaneq1})-(\ref{meaneq3}) 
the asymptotic behavior of the ground state solution for $x\gg\ell_B$ 
can be obtained as $v_{Iy}\sim (2\nu^{\prime} / h )( \left| m \right| / \rho_b )(q e^2 /x )$, $\rho_I - \rho_b / \left| m \right| \sim - (2{\nu^{\prime}}^2 / h^2 ) (\left| m \right| / \rho_b)$ 
$\left( \sum_J K_{IJ} M_J \right) (q e^2 / x^2 )$, and $\partial_x a_{I0}\sim 2 q e /x$ .
Then the mode obtained above decays algebraically perpendicular to the
edge, i.e., $\delta v_{Iy}^{(s)}\sim (2\nu^{\prime}/h)(\left| m \right|/\rho_b)
(e^2/x)$, $\delta \rho_I^{(s)}\sim -(2{\nu^{\prime}}^2/h^2)(\left| m \right|/\rho_b)$ 
$\left( \sum_J K_{IJ} M_J \right)(e^2/x^2)$ and $\partial_x \delta a^{(s)}_{I0}\sim (2e/x)$.
Then this mode can be identified as the 
edge magneto-plasmon because (i) it corresponds to the oscillation 
to the total 
charge $q$, i.e., the charge of each boson oscillates
in phase. (ii) The spatial extent 
of this charged mode into the bulk is much larger than 
the magnetic length $\ell_B$, and is insensitive to the details of the 
edge and the hierarchy structure of FQHL. This enables the classical
treatment as given by Volkov-Mikhailov \cite{volkov}.

Thus we have found one eigenmode of the system.
The next problem is if there are other charged modes which have the 
similar properties as that above, i.e., 
$\omega_k \propto k \ln k$ and the power-law decay.
To see this we analyze RPA equations
(\ref{rpaeq1})-(\ref{rpaeq3}) in the asymptotic region $x\gg\ell_B$. 
We are interested in the solutions 
with the  power law decay, i.e.,  
$\delta \rho_I \sim r_I/x^{\alpha} \cos (ky-\omega t)$, 
$\delta v_{Iy} \sim s_I/x^{\beta} \cos (ky-\omega t)$, 
$\partial_x \delta a_{I0} \sim t_I/x^{\gamma} \cos (ky-\omega t)$, where 
$r_I$, $s_I$, $t_I$ and 
$\alpha$, $\beta$, $\gamma$ are some unknown constants.
Putting these forms into eqs.(\ref{rpaeq1})-(\ref{rpaeq3}), we neglect
the terms which are the order of $k$ or $\omega$ and keep only the dominant
terms in the region $x\gg \ell_B$.
Then we obtain 
the exponents as $\alpha=2$, $\beta=\gamma=1$ and  
\begin{equation}
r_I=-\frac{\nu^{\prime}}{h}\sum_J K_{IJ} M_J s_J, 
\end{equation}
\noindent
\begin{equation}
s_I =\frac{\left| m \right|}{\rho_b} \frac{e\nu^{\prime}}{h} t_I,
\end{equation}
\noindent
\begin{equation}
t_I = -2e \int dx^{\prime} \delta \rho (x^{\prime}) .
\label{lasteq}
\end{equation}
\noindent
It should be noted that if the total charge 
$\int d x' \delta \rho(x')$
is finite, it coincides with the asymptotic behavior of the mode
obtained above.
From the uniqueness of the solution to the differential equations, 
it is concluded that the edge magneto-plasmon is single.
All the other modes are neutral in the sense that 
$\int d x' \delta \rho(x') = 0$, and hence decay faster than any
power-law, probably exponentially, within the magnetic length $\ell_B$.
This is because $\ell_B$ is the only length scale when the 
Coulomb interaction is irrelevant,
and the edge theory with short range interaction\cite{wen} \cite{nagaosa}
is applicable.
Note that this condition mentions only about the integral of the charge, and
generally the charge {\it density} of each neutral mode is not zero within the
length scale of $\ell_B$ from the edge.
This leaves the possibility to detect the neutral modes in terms of the 
electromagnetic measurements if the probe is localized within 
$\ell_B$  from the edge.

Now we study the other edge modes more explicitly. 
For this purpose we resort to the hydrodynamic treatment
regarding the FQHL as a completely incompressible fluid up to the 
edge \cite{giovanazzi}. 
This method is rather phenomenological and less exact, 
and can not reproduce the correct 
behavior perpendicular to the edge. However all the edge modes can be
derived in a simple way.
 We start from the following equation 
which can be obtained from eq.(\ref{meaneq1}) by taking the gradient.
\begin{eqnarray}
M_I \left[ \partial_t {\bf v}_I + \left( {\bf v}_I \cdot \nabla \right) {\bf v}_I \right]
 & = & \frac{h}{\nu^{\prime}} \sum_J K^{-1}_{IJ} \rho_J \left( {\bf v}_J - {\bf v}_I \right)
  \times \hat{e}_z \nonumber \\
 &   & -e {\bf E} - \frac{e}{c} {\bf v}_I \times {\bf B} + \nabla P_I,
\label{hmeaneq1}
\end{eqnarray}
\noindent
where 
\begin{eqnarray}
P_I &=& \frac{\hbar}{2M_I} \frac{1}{\sqrt{\rho_I}} \nabla^2 \sqrt{\rho_I}
+\lambda_I \left( \frac{\rho_b}{\left| m \right|} - \rho_I \right) 
\nonumber \\
& & - \int d^2 {\bf r}^{\prime} V \left( {\bf r}-{\bf r}^{\prime} \right) 
   \left[ \rho ({\bf r}^{\prime},t)-\rho_b \left( x^{\prime} \right) \right]
\label{pressure}
\end{eqnarray}
\noindent 
and $\hat{e}_z$ is unit vector of $z$-axis.
Now the approximation of the complete incompressibility is introduced.
Because the density fluctuation is zero \cite{giovanazzi} ,
$\nabla \times {\bf v}_I = 
(e / M_I c) \nabla \times ( {\bf A} + {\bf a}_I ) =0$ .
From this equation the velocity ${\bf v}_I$ is 
expressed by the scalar potential $\phi_I$ as
${\bf v}_I = \nabla \phi_I$.
In addition $\nabla \cdot {\bf v}_I = \nabla^2 \phi_I = 0$ 
from the equation of continuity . 
Then each ${\bf v}_I$ is expressed as 
${\bf v}_I = \phi_I^{(0)} \nabla  e^{iky-\left| k \right| x - i \omega t} $
with some constant $\phi_I^{(0)}$.
For $V({\bf r})$, we take the Coulomb interaction :
$V\left( {\bf r} \right)=e^2/ \left| {\bf r} \right|$.
Now we introduce an approximation for the pressure $P_I$.
We divide $P_I$ into two contributions. One is from the long range Coulomb
interaction, which corresponds to the third term 
in the rhs of eq. (\ref{pressure}). 
The other is the surface tension due to the short-range interaction
which roughly corresponds to the first and second terms.
We write the surface tension energy as 
$E_{S.T.}^{(I)}=(\lambda_I^{\prime}/2) \rho_b^2
\int dy \xi_I^2 \left( y \right) $ with $\xi_I$ being
the displacement of $\rho_I$ at $y$ 
and the pressure from it is given by
$(P_I)_{S.T.}= (\left| m \right| / \rho_b) (\delta E_{S.T}^{(I)} / \delta \xi_I)$.
The Coulomb interaction is taken into account by the 3-dimensional Coulomb 
integration \cite{giovanazzi} . 
The displacement $\xi_I$ is related to $\rho_I$ by the equation of continuity at
the boundary,
then we have from (\ref{hmeaneq1}):
\begin{eqnarray}
& & \omega \sum_J \frac{eB}{c}K_{IJ}^{-1} \phi_J^{(0)} 
 + k \lambda^{\prime}_I \rho_b \phi_I^{(0)} \nonumber \\
& & \hspace{15mm}  + 2 \rho_b e^2 k \log \frac{k_0}{\left| k \right|} \cdot \sum_J \phi_J^{(0)} = 0,
\label{hdisp}
\end{eqnarray}
\noindent
where we ignored the $\omega^2$-order term and $k_0^{-1}$ is some short range
cut off.
Note that $K^{-1}_{IJ} = \nu^{\prime} p - \nu^{\prime} \delta_{IJ}$, 
and define 
$a_I=-(eB/c) \nu^{\prime} \omega + \lambda_I^{\prime} \rho_b k$,
$b=(eB/c) p \nu^{\prime} \omega + 2e^2 \rho_b 
k \log (k_0/\left| k \right|)$ and
$A_{IJ}=b+a_I \delta_{IJ}$, eq.(\ref{hdisp}) becomes 
\begin{equation}
\sum_J A_{IJ} \phi_J^{(0)} = 0 .
\label{eigen}
\end{equation}
\noindent
Then the eigenvalues are obtained by $\det A = 0$.
This equation becomes
\begin{equation}
\left( \frac{eB}{c} p \nu^{\prime} \omega + 2e^2 \rho_b k \log \frac{k_0}{k} \right)
 \sum_{J=1}^{\left| m \right|} \frac{1}{\rho_b \lambda_I^{\prime} k - \frac{eB}{c} \nu^{\prime}\omega} +1=0.
\label{hheqdisp}
\end{equation}
\noindent
If the solution $\omega$ to eq.(\ref{hheqdisp}) is order of $k\ln k$,
we neglect the terms of order $k$ and obtain
\begin{equation}
\hbar \omega_k \sim - \nu \frac{e^2}{\pi} k \log \frac{k_0}{\left| k \right|}.
\label{hheqedgemagneto}
\end{equation}
\noindent
For the other solutions, it can be seen that $\omega\sim k$. 
Then $A_{IJ}\sim b$ and we get from eq.(\ref{eigen}), 
$\sum_{I=1}^{\left| m \right|} \phi_I^{(0)} = 0$.
Then we find these mode are neutral modes.
Especially in the case of $\nu=2/3$, the solutions to eq.(\ref{hheqdisp}) are
\begin{equation}
\hbar \omega_k \sim -\frac{2}{3} \frac{e^2}{\pi} k \log \frac{k_0}{\left| k \right|}
\end{equation}
\noindent 
and 
\begin{equation}
\hbar \omega_k \sim \frac{3}{2} \rho_b \ell_B^2 
    \left( \lambda_1^{\prime}+\lambda_2^{\prime} \right) k .
\end{equation}
\noindent 
The former mode is the edge magnetoplasmon and 
the latter mode is the neutral mode. The propagating directions of these
modes are opposite to each other.

Lastly we discuss the time-of-flight experiment in the $\nu=2/3$ FQHL 
in terms of the results obtained above.
As discussed below eq.(\ref{lasteq}), the neutral mode is not ``neutral''
locally within the length scale of $\ell_B$.
If one can detect such a local charge density,
the neutral mode propagating in the opposite direction
should show up. However in the experimental setup in ref.\cite{ashoori},
the charge is introduced to a wide region near the edge whose length scale
is typically $\sim 1 \mu m (\gg \ell_B)$. 
Hence the total charge $\int dx \delta \rho (x)$ integrated 
perpendicular to the edge is relevant, 
and only the edge magneto-plasmon 
should be detected as observed experimentally.

The authors thank S. Levit, X. G. Wen, M. Kohmoto, A. M. Finkel'stein
for useful discussions.

\end{document}